# Downstream: efficient cross-platform algorithms for fixed-capacity stream downsampling

**Connor Yang** [1,5], **Joey Wagner** [6,9], **Emily Dolson** [7,8,9], **Luis Zaman** [2,3,4], and **Matthew Andres Moreno** [2,3,4,5]

**1** Undergraduate Research Opportunities Program **2** Department of Ecology and Evolutionary Biology **3** Center for the Study of Complex Systems **4** Michigan Institute for Data and AI in Society **5** University of Michigan, Ann Arbor, United States **6** Professorial Assistantship Program **7** Department of Computer Science and Engineering **8** Program in Ecology, Evolution, and Behavior **9** Michigan State University, East Lansing, United States

## Summary

Due to ongoing accrual over long durations, a defining characteristic of real-world data streams is the requirement for rolling, often real-time, mechanisms to coarsen or summarize stream history. One common data structure for this purpose is the ring buffer, which maintains a running downsample comprising most recent stream data. In some downsampling scenarios, however, it can instead be necessary to maintain data items spanning the entirety of elapsed stream history. Fortunately, approaches generalizing the ring buffer mechanism have been devised to support alternate downsample compositions, while maintaining the ring buffer's update efciency and optimal use of memory capacity ( Gunther, 2014; Moreno et al., 2024). The Downstream library implements algorithms supporting three such downsampling generalizations: (1) "steady," which curates data evenly spaced across the stream history; (2) "stretched," which prioritizes older data; and (3) "tilted," which prioritizes recent data. To enable a broad spectrum of applications ranging from embedded devices to high-performance computing nodes and AI/ML hardware accelerators, Downstream supports multiple programming languages, including C++, Rust, Python, Zig, and the Cerebras Software Language. For seamless interoperation, the library incorporates distribution through multiple packaging frameworks, extensive cross-implementation testing, and cross-implementation documentation.

## Statement of Need

Efficient data stream processing is crucial in modern computing systems, where workloads of continuous, high-volume data have become more prevalent (Cordeiro & Gama, 2016). Applications of data stream processing include sensor networks (Elnahrawy, 2003), distributed big-data processing (He et al., 2010), real-time network trafc analysis ( Johnson et al., 2005; Muthukrishnan, 2005), systems log management (Fischer et al., 2012), fraud monitoring (U & Babu, 2016), trading in financial markets (Agarwal et al., 2009), environmental monitoring (Hill et al., 2009), and astronomical surveys (Graham et al., 2012), which all generate data at rates that exceed practical storage capacity, while requiring analysis across varying time horizons.

Within the broader ecosystem of tools for data stream processing, Downstream targets, in specifc, use cases that require best-effort downsampling within fxed memory capacity. Within this domain, Downstream especially benefts scenarios involving:

1. real-time operations, due to support for $\mathcal{O}(1)$ data processing;

2. need for compact memory layout with minimal overhead, especially where individual data items are small (e.g., single bits or bytes) relative to bookkeeping metadata or where curated downsamples are frequently copied/transmitted; and/or
3. support for SIMD acceleration (e.g., ARM SVE, x86 AVX, GPU, etc.), due to branchless structure of underlying algorithms.

As such, Downstream is well-suited to emerging AI/ML hardware accelerator platforms, such as Cerebras Systems' Wafer-Scale Engine (WSE) (Lie, 2023), Graphcore's Intelligence Processing Unit (Gepner et al., 2024), Tenstorrent's Tensix processors (Vasiljevic et al., 2021), and Groq's GropChip (Abts et al., 2022). Pursuing an aggressive scale-out paradigm, these platforms bring hundreds, thousands, or — in the case of the Wafer-Scale Engine — hundreds of thousands of processing elements to bear on a single chip. As a design trade-off, however, on-device memory available per processor on these platforms is generally scarce. For instance, although the Cerebras WSE-2 supplies 40 gigabytes of on-chip memory in total, split between processor elements this amounts to less than 50 kilobytes each (Lie, 2023). Indeed, a key use case motivating the Downstream library has been in managing data for agent-based evolution experiments conducted on the WSE platform, a topic discussed further in "Projects Using This Software."

## Approach

Algorithms in Downstream consist of two components:

1. *site selection*, which controls ongoing runtime downsample curation, and
2. *site lookup*, which identifes stream arrival indices (i.e., timepoints) for stored data.

Downstream's runtime data management applies a minimalistic strategy. First, a fixed-capacity working buffer is assumed, with user-defned size. Second, stored data remains fixed in place without subsequent editing or relocation. Operations on ingested items are therefore limited to storing, discarding (i.e., without storage), or overwriting previously stored data. Hence, downsample curation is wholly determined by "site selection" — i.e., the placement of each ingested data item. This scheme, in essence, represents a generalized ring buffer, and — as such — provides compact, efficient processing and storage (Gunther, 2014).

Figure 1 illustrates Downstream's single-operation "site selection" approach. At any point, but typically in a postprocessing step, a corresponding "site lookup" procedure can calculate stored items' arrival index, allowing this metadata to be omitted in storage.

In practice, typical nuts-and-bolts steps for end-user code are thus: (1) initialize a fixed-size buffer with desired capacity ($S$), (2) maintain a count $T$ of elapsed stream depth, and (3) use $T$ and $S$ to call the site selection method of a chosen Downstream algorithm to place each arriving stream item in the buffer (or discard it). From this point forward, steps in using or analyzing stored data will vary by use case. Among a variety of supported possibilities, one simple workfow would be to (1) dump memory segments comprising counter $T$ and stored buffer content as hexidecimal strings in a tabular data fle (e.g., Parquet, CSV, etc.) then (2) using Downstream CLI, explode as long format (i.e., one row per data item) with corresponding data item lookups (i.e., stream arrival index).

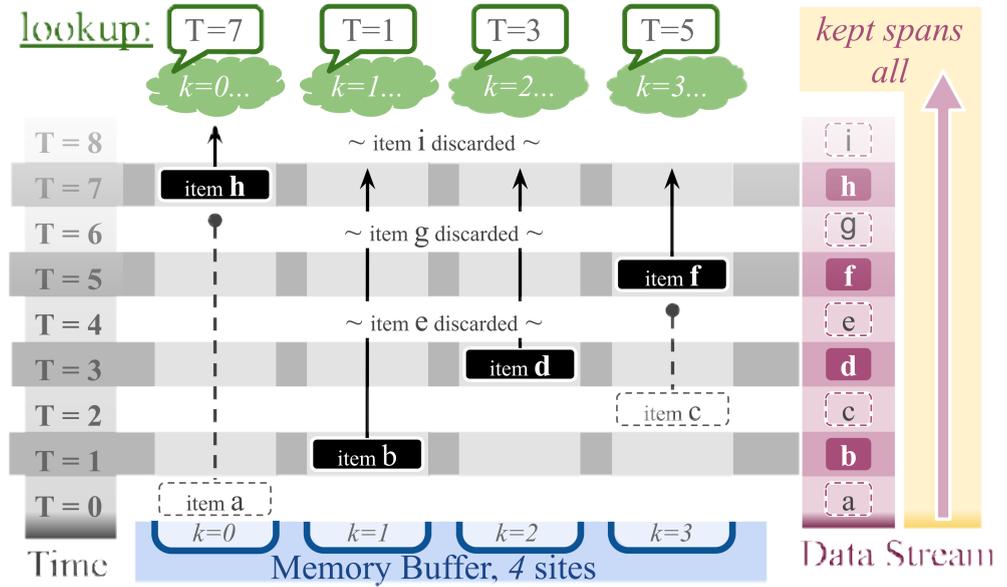

**Figure 1:** Schematic illustration of a Downstream site selection algorithm operating on a data stream with a fixed-capacity memory buffer ($S = 4$ sites). Each new item in the data stream (right) arrives over time ($T \in [0, 8]$), and is either stored in the memory buffer or discarded. Storage decisions are made independently at each timestep, with each item mapped to one of the $S$ memory sites ($k \in 0, 1, 2, 3$). Previously stored items may be overwritten. This example illustrates curation of a steady-spaced downsample. Green boxes (top) depicts lookup operation to calculate stream indices $L(T)$ of the most recently stored data at each site at time $T = 8$.

## Features

Downstream provides algorithms for curating stream downsample density according to three primary temporal distributions: steady, stretched, and tilted.

The **steady algorithm** maintains uniform spacing between retained items. This approach is best suited for applications in which it is important to maintain data from all time periods, such as for trend analysis in long-term monitoring systems. In addition to an approach proposed in (Moreno et al., 2024), Downstream includes Python implementation of the "compressing ring buffer" approach for steady curation developed by (Gunther, 2014).

The **stretched algorithm** prioritizes older data while maintaining recent context, focusing on preserving the origins of the stream. Specifcally, the density of retained data is thinned proportionally to depth in the stream. This approach suits applications where detailed understanding of initial conditions is critical.

The **tilted algorithm** prioritizes recent information over older data. Specifcally, the density of retained data is thinned proportionally to age. This makes it well-suited for monitoring and alerting systems where recent trends are most relevant, but historical context still provides valuable perspective — such as in real-time monitoring systems (Tabassum & Gama, 2016), where recent data carries more operational signifcance than older data (Aggarwal, 2006).

**Table 1:** Comparison of core Downstream algorithms.

| Algorithm Name | Distribution of Retained Data | Example Use Case |
| --- | --- | --- |
| Steady | Evenly distributed | Long-term systems monitoring |

| Algorithm Name | Distribution of Retained Data | Example Use Case |
| --- | --- | --- |
| Tilted | Favors recent data | Ancestry markers in evolutionary simulations |
| Stretched | Favors older data | Real-time alert monitoring |

To support diverse end-user integrations, Downstream has been implemented across five programming languages: C++, Rust, Python, Zig, and the Cerebras Software Language (CSL). We have organized each implementation as a standalone branch within the library's git repository.

For all implementations, we provide:

1. Steady, stretched, and tilted site selection methods.
2. API documentation, to demonstrate function signatures and semantics.
3. Installation instructions, through standard package managers where supported (e.g., Python's PyPI, Cargo's Rust). C++ code is provided as a header-only library.
4. Extensive validation tests, ensuring complete interchangeability and exact compatibility between platforms (e.g.,for separate data collection and analysis steps).

On an as-needed basis, implementations of additional hybrid algorithms are provided, which split buffer space between multiple temporal distributions. Support for high-throughput bulk lookup operations is implemented in Python, with both CLI- and library-based interfaces available. A Python-based CLI is also provided for validation testing, facilitating the development of additional implementations for new languages or platforms.

## Empirical Scaling Benchmark

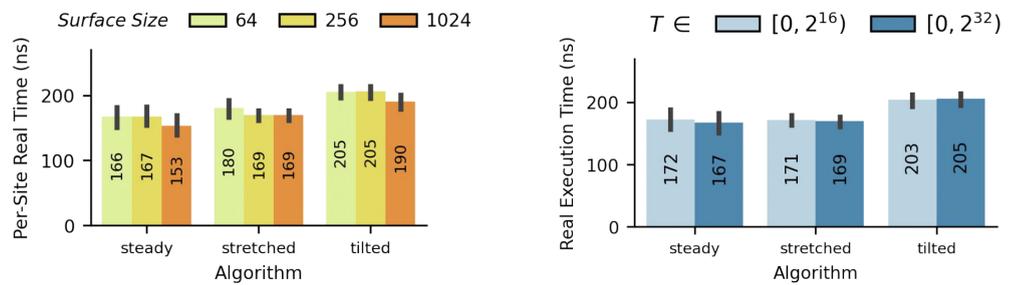

Figure 2: Execution time of Downstream site selection algorithms across varying runtime environments. (Left) Per-site real execution time across diferent surface sizes ( $S \in \{64, 256, 1024\}$), representing the size of the buffer. (Right) Real execution time across diferent time ranges ( $T \in [0, 2^{16})$ vs. $[0, 2^{32})$). Bars show bootstrap 95% confdence intervals.

A key goal of Downstream is efficient scaling to large buffer sizes and deep stream durations. To test the library's performance, we conducted empirical benchmarking trials of Python site selection methods. Shown in Figure 2, we observed consistent execution time across both buffer size and stream depth (i.e., number of data points processed). Statistical analysis detected a signifcant efect ( $\alpha = 0.05$; Kruskal–Wallis; $n = 30$), but we did not observe evidence of worse performance with increasing surface size — in fact, larger surface sizes were associated with lower execution times. As shown in Figure 2, this result could potentially reflect efficiency gains from larger batch sizes.

### Projects Using the Software

A motivating use case for the Downstream library is in supporting work on *hereditary stratigraphy*, a decentralized, approximate phylogeny tracking method for large-scale, parallel and distributed agent-based evolution simulations. In this use case, phylogenetic history (i.e., a population's ancestry tree) is estimated from Downstream-curated data embedded in agent genomes. This data comprises a running downsample of randomly-generated checkpoint values, generated and appended for each generation elapsed. By comparing these checkpoint values, it is possible to estimate when the lineages of extant genomes diverged (Moreno et al., 2022a). Notably, this use case can make use of both steady and tilted distributions (Moreno et al., 2025), and depends on post-hoc stream index lookup to identify the generational timing of inferred phylogenetic events.

To this end, Downstream serves as a key dependency for the library implementing hereditary stratigraphy methodology, *hstrat* (Moreno et al., 2022b). In recent work with *hstrat*, the CSL Downstream implementation has been applied to support phylogeny tracking in massively distributed, agent-based evolution simulations conducted on the 850,000-processor WSE platform (Moreno et al., 2024). In other forthcoming work employing WSE-based simulations of hypermutator evolution, Downstream has also been used to collect time series data leading up to in-simulation extinction events.

In both examples, Downstream provides a mechanism for best-efort trade-offs between runtime efciency and data quality, wherein a considered amount of precision (chosen based on experimental objectives) is exchanged for memory savings, dynamic fexibility, and — ultimately — increased scalability. We anticipate this pattern continuing as a recurring theme in further applications of the library.

### Related Software

Several notable projects provide data stream processing functionality related to Downstream.

The most similar piece of software to Downstream is Gunther's work on compressing circular buffers (Gunther, 2014). This approach exploits modular arithmetic as the basis for a ring buffer generalization with steady-spaced coarsening behavior. Included software, written in Java, notably supports additional aggregation algorithms (e.g., averaging, summing, extrema, etc.) in addition to sampling, as is the focus in Downstream. Downstream extends beyond (Gunther, 2014), however, in enabling support for stretched and tilted downsampling, as well as cross-language support and high-throughput lookup decoding.

Reservoir sampling approaches provide representative samples of data streams, but lack deterministic temporal distribution guarantees and metadata-free stream arrival index attribution provided by Downstream (Aggarwal, 2006; Hentschel et al., 2018).

Apache Flink and Spark Streaming are general-purpose distributed computing frameworks for stream processing, which focus on distributed computation over massive data streams rather than data downsampling (Carbone et al., 2015; Salloum et al., 2016).

InfuxDB and TimescaleDB are time-series databases that ofer storage solutions for continuous data in applications like IoT devices, but typically require open-ended storage capacity or downsampling strategies fundamentally diferent from Downstream's algorithms ( Naqvi et al., 2017; Stefancova, 2018).

### Future Work

As originally proposed (Moreno et al., 2024), Downstream's stretched and tilted algorithms only support stream sizes up to $2^S - 2$ items (where $S$ is the buffer size). Preliminary work, included in recent releases of Downstream, is underway to explore extensions beyond this point. Work is also underway on more comprehensive cross-platform benchmarks.

More generally, we plan to continue developing library features — e.g., extending partially-supported functionality across language branches, supporting additional programming languages, etc. — on an as-needed basis, and welcome user requests or contributions to this end.

## Acknowledgements

Thank you to Vivaan Singhvi for contributing supplemental material, including benchmarks and additional implementations, to the Downstream and hstrat libraries. This research was supported by the University of Michigan through the Undergraduate Research Opportunities Program, by Michigan State University through computational resources provided by the Institute for Cyber-Enabled Research, and by the Eric and Wendy Schmidt AI in Science Postdoctoral Fellowship, a Schmidt Sciences program.

This material is based upon work supported by the U.S. Department of Energy, Ofce of Science, Ofce of Advanced Scientifc Computing Research (ASCR), under Award Number DE-SC0025634. This report was prepared as an account of work sponsored by an agency of the United States Government. Neither the United States Government nor any agency thereof, nor any of their employees, makes any warranty, express or implied, or assumes any legal liability or responsibility for the accuracy, completeness, or usefulness of any information, apparatus, product, or process disclosed, or represents that its use would not infringe privately owned rights. Reference herein to any specifc commercial product, process, or service by trade name, trademark, manufacturer, or otherwise does not necessarily constitute or imply its endorsement, recommendation, or favoring by the United States Government or any agency thereof. The views and opinions of authors expressed herein do not necessarily state or refect those of the United States Government or any agency thereof.